\documentstyle[psfig,twoside,fleqn,espcrc2]{article}

\newcommand{\AmS}{{\protect\the\textfont2
  A\kern-.1667em\lower.5ex\hbox{M}\kern-.125emS}}

\title{Further Evidence of a Smooth Phase in 4D Simplicial Quantum \\ Gravity
\thanks{presented by H.S.Egawa}}

\author{H.S.Egawa
        \address{Department of Physics, Tokai University, 
        Hiratsuka, Kanagawa 259-1292, Japan}
        $^{,}$
        \address{Theory Division, Institute of Particle 
        and Nuclear Studies, KEK, High Energy Accelerator 
        Research Organization, Tsukuba, Ibaraki 305-0801, Japan}
        ,
        A.Fujitsu
        \address{Information Systems and Technology Center, 
        The University of Aizu, Tsuruga, Ikki-machi, Aizu-Wakamatsu, 
        Fukushima 965-8580, Japan}
        ,
        S.Horata$^{\,\, {\rm a}}$
        ,
        N.Tsuda$^{\,\, {\rm b}}$
        \thanks{supported by Research Fellowship of the Japan 
        Society for the Promotion of Science for Young Scientists.}
        and 
        T.Yukawa 
        \address{Coordination Center for Research and Education, 
        The Graduate University for Advanced Studies, 
        Hayama, Miura, Kanagawa 240-0193, Japan}
        $^{\! , \,\,{\rm b}}$}
\begin{document}

\begin{abstract}
Four-dimensional (4D) simplicial quantum gravity coupled to U(1) 
gauge fields has been studied using Monte-Carlo simulations.
A negative string susceptibility exponent is observed beyond the 
phase-transition point, even if the number of vector fields ($N_{V}$) 
is $1$. 
We find a scaling relation of the boundary volume distributions 
in this new phase.
This scaling relation suggests a fractal structure similar to that of 
2D quantum gravity.
Furthermore, evidence of a branched polymer-like structure is 
suggested far into the weak-coupling region, even for $N_{V} > 1$.
As a result, we propose new phase structures and discuss the 
possibility of taking the continuum limit in a certain region 
between the crumpled and branched polymer phases. 
\end{abstract}
\maketitle 
\section{Introduction} 
The development of simplicial quantum gravity started with the 2D case. 
Recently, the phase structure for 4D pure simplicial quantum gravity 
has been intensely investigated as a first step. 
In 4D pure gravity, two distinct phases are known.
For small values of the bare gravitational coupling constant the 
phase is the so-called elongated phase, which has the characteristics 
of a branched polymer.
For large values of the bare gravitational coupling constant the phase is 
the so-called crumpled phase.
Numerically, the phase transition between the two phases has been shown 
to be 1st order. 
As a result, it is difficult to construct a continuum theory. 
Our second step is to investigate an extended model of 4D quantum gravity. 
From calculations in ref.\cite{AMM} we have tried introducing vector fields.
Actually, we have treated pure gravity coupled to U(1) gauge fields 
and have considered the possibility of taking a continuum limit. 
In order to investigate the phase structures, we mainly measured the 
string susceptibility exponent ($\gamma_{st}$) using the Minbu method \cite{JM} 
and the boundary volume distribution \cite{EHITY}. 
The aim of this article is to discuss the new phase (smooth phase) in 
$4$D simplicial quantum gravity coupled to gauge fields. 
\vspace{-1mm}
%
\section{Models with Gauge Fields}
We start with the Euclidean Einstein-Hilbert action in $4$D for pure gravity: 
\begin{equation}
S_{EH}[\Lambda,G] = \displaystyle{\int} d^{4}x \sqrt{g}(\Lambda-\frac{1}{G}R), 
\end{equation}
where $\Lambda$ is the cosmological constant and $G$ is Newton's constant.
We use discretize action for pure gravity, 
$
S_{P}[\kappa_{2},\kappa_{4}] = \kappa_{4}N_{4}-\kappa_{2}N_{2},
$
where $\kappa_{2} \sim \frac{1}{G}$, $\kappa_{4}$ is related to $\Lambda$ and 
$N_{i}$ is the number of $i$-simplexes.
We use the plaquette action for U(1) gauge fields \cite{BBKPTT}, 
\begin{equation}
S_{G}=\sum_{t_{ijk}}o(t_{ijk})[A(L_{ij})+A(L_{jk})+A(L_{ki})]^{2},
\end{equation}
where $L_{ij}$ denotes a link with vertices $i$ and $j$, $t_{ijk}$ denotes 
a triangle with vertices $i$, $j$ and $k$, $A(L_{ij})$ denotes the $U(1)$ 
gauge field on a link $L_{ij}$ and $o(t_{ijk})$ denotes the number of 
simplexes sharing triangle $t_{ijk}$.
The total action of pure gravity with $U(1)$ gauge fields is $S=S_{P}+S_{G}$. 
We consider a partition function,
$
Z(\kappa_{2},\kappa_{4}) = \sum_{T}W(T) \int \prod_{L \in T} 
dA(L)e^{-S_{P}-S_{G}},
$
where $W(T)$ is the symmetry factor. 
We sum over all 4D simplicial triangulations ($T$) in order to carry out 
a path integral about the metric. 
Here, we fix the topology with $S^{4}$.
%
\section{Numerical Results}
%
\begin{figure}
\centerline{\psfig{file=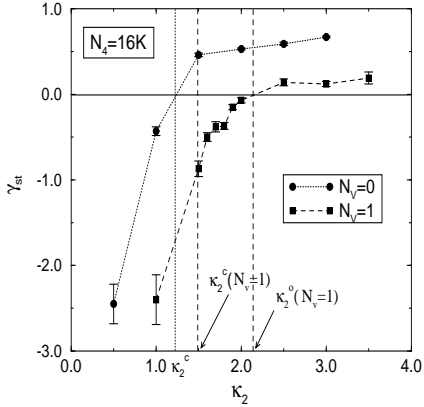,height=6cm,width=6cm}} 
\vspace{-13mm}
\caption
{
String susceptibility exponents ($\gamma_{st}$) plotted versus the 
coupling constant ($\kappa_{2}$) for $N_{v}=0$ and $1$.
}
\label{fig:gamma}
\vspace{-0.7cm}
\end{figure}
%
In this section we report on two numerical observations: the $\gamma_{st}$ 
and the boundary volume distributions.
The $\gamma_{st}$ is defined by the asymptotic form of the partition 
function,
$
Z(V_{4}) \sim V_{4}^{\gamma_{st} - 3} e^{\mu V_{4}},
$
where $V_{4}$ denotes the $4$D volume.
In Fig.1 we plot $\gamma_{st}$ for various numbers of gauge fields 
versus $\kappa_{2}$ with volume $N_{4}=16K$.
What is important in the $N_{V}=1$ case is that the usual phase-transition 
point ($\kappa_{2}^{c}$) is different from another transition point 
($\kappa_{2}^{o}$) which separates the $\gamma_{st} < 0$ region from the 
$\gamma_{st} > 0$ region and $\gamma_{st}$ becomes negative at the 
phase-transition point $\kappa_{2}^{c}$.
This fact leads to the definition of a new smooth phase.
The new smooth phase is defined by an intermediate region between these 
two transition points, $\kappa_{2}^{c}$ and $\kappa_{2}^{o}$.
In the pure-gravity case it is clear that $\kappa_{2}^{c} \approx 
\kappa_{2}^{o}$, and thus there is no evidence for the existence of a new 
smooth phase. 
On the other hand, in the case of $N_{V}=1$ with $N_{4}=16K$, we observe 
the $\gamma_{st} < 0$ region beyond the usual phase-transition point 
($\kappa_{2}^{c}$).
We also observe a very obscure transition from $\gamma_{st} < 0$ to 
$\gamma_{st} > 0$ at $\kappa_{2}^{o}$ (see in Fig.1).
This obscure transition is very similar to that of $c=1$ in 2D quantum 
gravity.
In 2D the $c=1$ barrier is well known as an obscure transition from 
the fractal phase ($c \leq 1$ and $\gamma < 0$) to the branched polymer 
phase ($c > 1$ and $\gamma > 0$).
In the $N_{V}=1$ case we observe a smooth phase which is separated from the 
crumpled phase by $\kappa_{2}^{c}$, and observe the branched polymer phase 
which is separated from the smooth phase by $\kappa_{2}^{o}$.
In order to investigate statistical structures of these three phases we 
have observed the boundary volume distributions ($\rho$) in $4$D Euclidian 
space-time using the concept of geodesic distances.
In order to discuss the universality of the scaling relations, we 
assume that $D^{-\alpha} \cdot \rho(V,D)$ is a function of a scaling 
variable, $x=V/D^{\alpha}$ \cite{EHITY}.
Here, $V$ denotes the volume of the boundary and $D$ is the geodesic distance.
This assumption has been justified in $2$D \cite{KKMW}.
The scaling parameter $\alpha = d_{f}-1$ is also defined in the same manner 
as in ref.\cite{EHITY}.
Here, $d_{f}$ denotes the fractal dimension.
In Fig.$2$ we plot the boundary volume distributions for various geodesic 
distances for $N_{V}=1$ with $N_{4}=16K$ in the smooth phase 
($\kappa_{2}=1.7$).
The distributions at different distances show excellent agreement with 
each other.
It is clear that the $4$D simplicial manifold becomes fractal in the sense 
that sections of the manifold at different distances from a given $4$-simplex 
look exactly the same after a proper rescaling of the boundary volume.
Furthermore, the shape of this scaling function is very similar to that of 
the $2$D case \cite{KKMW,TY}.
The best account for this excellent agreement in the $4$D case can be found 
in the dominance of a conformal mode and a fractal property.
It seems reasonable to suppose that this new smooth phase has a similar 
fractal structure to that of the $2$D fractal surface, and has the 
possibility of taking a continuum limit.
We have also investigated the boundary volume distribution in the crumpled and 
the branched polymer phases.
In the crumpled phase we find that one mother universe is dominant.
On the other hand, in the branched polymer phase we have no evidence for 
the existence of a mother universe.
There is one further observation that we must not ignore in the region 
$\kappa_{2} > \kappa_{2}^{o}$.
The number of nodes of the manifolds is very close to its upper kinematic 
bound, $\frac{N_{0}}{N_{4}} \approx \frac{1}{4}$.
This upper kinematic bound of the simplexes serves as evidence of a 
branched polymer.
The phase transition at $\kappa_{2}^{c}$ becomes softer the larger $N_{V}$ 
becomes.
Actually, in the $N_{V}=3$ case a single peak in the node susceptibility is 
reported by the authors in ref.\cite{BBKPTT}.
Unfortunately, even in the $N_{V}=1$ case we have observed a discontinuity 
at the critical point $\kappa_{2}^{c}$, which is consistent with ref.\cite{BBKPTT}.
%
\begin{figure}
\centerline{\psfig{file=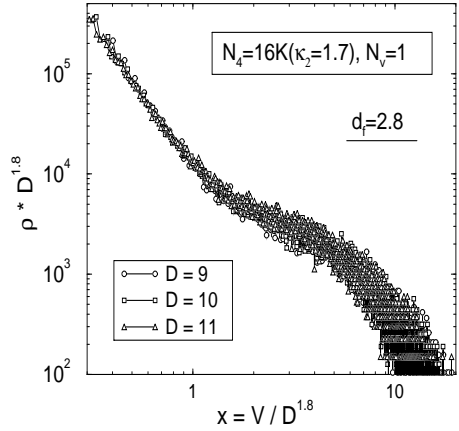,height=6.5cm,width=6.5cm}} 
\vspace{-13mm}
\caption
{
Boundary volume distributions plotted versus the scaling variable ($x$) 
with $N_{4}=16K$ (in the smooth phase: $\kappa_{2}=1.7$) for $N_{V}=1$ 
using $log-log$ scales. 
}
\vspace{-0.6cm}
\label{fig:Boundary_Volume}
\end{figure}
%
\begin{figure}
\centerline{\psfig{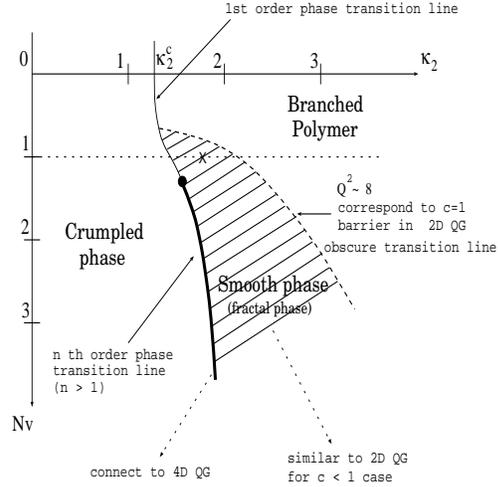}} 
\vspace{-1cm}
\caption
{
Rough sketch of the phase diagram.
The $N_{V}=1$ case has been intensively investigated, and we have 
obtained the scaling of the boundary volume distributions in Fig.2 
at the point indicated by the cross.
}
\vspace{-0.5cm}
\label{fig:Phase_4DQG_G}
\end{figure}
%
\vspace{-1mm}
%
\section{Summary and Discussions}
Let us summarize the main points made in the previous section.
In Fig.3 we show a rough sketch of the phase diagram of 4D simplicial 
gravity.
We have three phases in this parameter space: a crumpled phase, a smooth 
phase (shaded portion) and a branched polymer.
The thin line denotes a discontinuous phase-transition line which is 
known in pure gravity; the a thick line denotes a smooth 
phase-transition line.
In the smooth phase with $N_{4}=16K$ ($\kappa_{2}=1.7$) and $N_{v}=1$ we 
obtained $\gamma_{st}=-0.38(5)$, $d_{f}=2.8(5)$ and a good scaling 
relation of the boundary volume distributions with the scaling variable 
$x=V/D^{d_{f}-1}$.
The scaling structure of this smooth phase is similar to that of a 2D 
random (fractal) surface.
%
It suggests the existence of a new smooth phase in 4D simplicial gravity.
We obtained an obscure transition line (a broken line in  Fig.3), and 
suggest that the obscure transition corresponds to the $c=1$ barrier in 
2D quantum gravity.
The existence of genuine 4D quantum gravity on the critical point 
$\kappa_{2}^{c}$ remains a matter for discussion.

We are indebted to Masaki Yasue and Tsunenori Suzuki for their help.

\end{document}